\documentclass{elsart5p}

\usepackage{graphicx}
\usepackage{amssymb}
\usepackage{cite}		
\begin{document}

\begin{frontmatter}

\title{Ground-state degeneracy and low-temperature thermodynamics of 
correlated electrons on highly frustrated lattices}

\author[Goe]{A. Honecker\corauthref{AH}\ead{ahoneck@uni-goettingen.de}},
\author[Lviv]{O. Derzhko},
\author[MD]{J. Richter}

\address[Goe]{Institut f\"ur Theoretische Physik,
          Georg-August-Universit\"at G\"ottingen,
	  Friedrich-Hund-Platz 1,
          37077 G\"ottingen, Germany}
\address[Lviv]{Institute for Condensed Matter Physics,
          National Academy of Sciences of Ukraine,
          L'viv-11, 79011, Ukraine}
\address[MD]{Institut f\"ur Theoretische Physik,
          Universit\"at Magdeburg,
          P.O. Box 4120, 39016 Magdeburg, Germany}

\corauth[AH]{Corresponding author. Tel/Fax:
+49 551 397693/+49 551 399263}

\begin{abstract}

Highly frustrated lattices yield a completely flat lowest single-electron band.
Remarkably, exact many-body ground states can 
be constructed for the repulsive Hubbard model and the $t-J$ model by
filling this flat band with localized electron states.
This construction leads to a macroscopic ground-state degeneracy.
We discuss how to compute these ground-state degeneracies
for a certain class of models, including in particular the sawtooth chain. 
Furthermore, we discuss generic consequences for low-temperature 
thermodynamic properties, like the appearance of a low-temperature
peak in the specific heat.
Finally, we present complementary numerical results obtained by 
exact diagonalization.

\end{abstract}

\date{11 August 2008; revised 6 March 2009}


\end{frontmatter}

\section{Introduction}

It is a rare event that one can make exact statements about a
strongly correlated electron system, especially in dimensions
bigger than one. Competing interactions usually render the problem
even more difficult, but sometimes they are also helpful for the
analysis. In particular, on highly frustrated lattices it is
possible to construct a macroscopic number of ground states (GSs) using
localized single-particle excitations as building blocks.
On the one hand, such a construction has been applied to the so-called
flat-band Hubbard models in order to show that they exhibit fully
saturated ferromagnetism for suitable electron fillings
(see, e.g., \cite{mielke,tasaki,tasaki98,IKWO98,NGK}).
On the other hand, so-called localized magnon states have been found
to yield exact many-body GSs for highly frustrated quantum
antiferromagnets in high magnetic fields
\cite{jumpMol,jump,RSH04,RSHSS04}. Furthermore, it has been shown
that the localized-magnon GSs have important consequences
for the low-temperature magneto-thermodynamic properties like an
enhanced magnetocaloric effect close to the saturation field
\cite{ZhiHo,DeRi,ZhiTsu,RDK06,cooling2,review07,cmt31}.
Surprisingly, the corresponding thermodynamics of flat-band Hubbard models
did not seem to have been investigated despite the long history of
these models \cite{mielke,tasaki,tasaki98}. This
motivated us to start studying the thermodynamic properties of correlated
electron systems on highly frustrated lattices
\cite{cmt31,honecker_richter,sawtooth-tJ,sawtooth07,monomer08}.
Here we summarize some aspects of the construction of localized-electron
GSs as well as their consequences for the low-temperature
thermodynamics and present some complementary numerical results.

\section{Models with localized electron states}

We will discuss two classes of models of correlated electrons. Firstly, we
consider the $N$-site Hubbard Hamiltonian
\begin{eqnarray}
H&=&\sum_{\sigma=\uparrow,\downarrow}
\sum_{\langle i,j \rangle}t_{i,j}
\left(c_{i,\sigma}^{\dagger}c_{j,\sigma}
+c_{j,\sigma}^{\dagger}c_{i,\sigma}\right)
+U\sum_{i}n_{i,\uparrow}n_{i,\downarrow}
 \nonumber \\
&&  +\mu \sum_{i=1}^N n_{i} \, ,
\label{eq:Hub}
\end{eqnarray} 
where $i$ denotes the lattice sites,
$\langle i,j\rangle$ are pairs of nearest neighbors,
the $c_{i,\sigma}^{(\dagger)}$ 
are the usual fermion operators,
$n_{i,\sigma} = c_{i,\sigma}^{\dagger} c_{i,\sigma}$,
$n_{i} = n_{i,\uparrow}+n_{i,\downarrow}$.
$U\ge 0$ is the on-site Coulomb repulsion.
Note that we have chosen \cite{honecker_richter} non-standard
sign conventions for the hopping integrals $t_{i,j}$ and the chemical
potential $\mu$ in order to emphasize the analogy with the antiferromagnetic
Heisenberg model in a magnetic field.

Secondly, we consider the $t-J$ model 
with Hamiltonian
\begin{eqnarray}
H &=& \sum_{\sigma=\uparrow,\downarrow}
\sum_{\langle i, j\rangle} t_{i,j} \, P\, \left(
{c}^{\dagger}_{i,\sigma} {c}_{j,\sigma}
 + {c}^{\dagger}_{j,\sigma} {c}_{i,\sigma}\right)\,P \nonumber \\
&& + \sum_{\langle i, j\rangle} J_{i,j}
 \left(\vec{S}_i \cdot \vec{S}_j - {1 \over 4} \, n_i \, n_j \right)
+ \mu \sum_{i=1}^N {n}_{i} \, .
\label{eq:tJ}
\end{eqnarray}
Here $P$ is the projector which eliminates doubly occupied sites
and $\vec{S}_i$ are spin-1/2
operators acting on a singly occupied site $i$.
The $t-J$ model (\ref{eq:tJ}) arises as the large-$U$ limit
of the Hubbard model (\ref{eq:Hub}), yielding the relation
$J_{i,j} = 4\,t_{i,j}^2/U$ up to second order in the hopping
integrals $t_{i,j}$.
However, the implied relation between the
$J_{i,j}$ is not important for our purposes.
We will therefore ignore it 
and choose $J_{i,j}=J$. 
Conversely,
one could choose a site-dependent Coulomb repulsion $U_i > 0$ in the
Hubbard model (\ref{eq:Hub})
without affecting any of our main conclusions.

\begin{figure}[!t]
\begin{center}
\includegraphics[width=0.8\columnwidth]{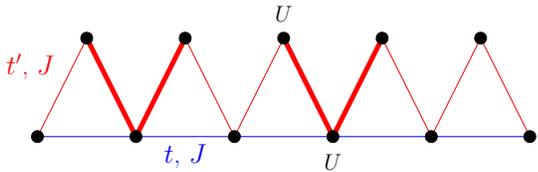}
\end{center}
\caption
{
The sawtooth chain. Filled circles show electron sites.
Two trapping cells occupied by localized electron states are
indicated by bold lines.
}
\label{fig:sawtooth}
\end{figure}

We would like to emphasize that our main conclusions
apply to any highly frustrated lattice.
However, for the sake of concreteness,
we will focus on the sawtooth chain sketched in Fig.~\ref{fig:sawtooth}.
The hopping integrals $t_{i,j}$ are $t$ along the base line and $t'$ 
along the zigzag-line, respectively.
Periodic boundary conditions are imposed along the chain direction.

The single-electron problem is solved as usually by introducing a
momentum $k$.
On the sawtooth lattice one finds two branches whose dispersions read
\begin{equation}
\varepsilon_{\pm}(k) = t\, \cos{k} \pm
\sqrt{t^2 \, \cos^2{k} + 2\, t'^2 \, \cos{k} + 2 \, t'^2}
+ \mu \, .
\label{eq:sawDisp}
\end{equation}
Remarkably, the lower branch $\varepsilon_{-}(k)$
becomes completely flat, {\it i.e.}, $k$-independent
for $t' = \sqrt{2}\,t$.
The sawtooth Hubbard model (\ref{eq:Hub})
with this choice of hopping integrals is a particular case of Tasaki's
model which  exhibits saturated ferromagnetism
for a half-filled flat band,
{\it i.e.}, when the number of electrons is $n=N/2$ \cite{tasaki,tasaki98}.
It is common practice to
introduce on-site energies for Tasaki's model \cite{tasaki,tasaki98} such that
the flat-band condition $t' = \sqrt{2}\,t$ can be replaced by a condition
for the on-site energies (see also \cite{IKWO98,NGK}).
However, since this generalization does not change any of the fundamental
physics, we will not pursue this either here.

There are many frustrated lattices which yield a lowest
completely flat single-electron band, including popular lattices like
the kagome and pyrochlore lattices in two and three dimensions, respectively.
In fact, a completely flat lowest single-electron band can be taken as the
defining property of a highly frustrated lattice.

Given such a flat band of excitations, one can transform back to real space
and localize the excitations in a finite region. Such `localized electron
excitations' live in trapping cells which for the sawtooth chain are
the valleys formed by three neighboring sites (see bold lines in
Fig.~\ref{fig:sawtooth}). Typically, each site adjacent to a trapping
cell is connected to several sites of the trapping cell such that destructive
quantum interference between the different paths prevents escape
of the electron. Now one can occupy each trapping cell
independently by electrons with the two different spin projections.
For $U=0$, this yields many-electron GSs of the Hubbard model
(\ref{eq:Hub}) by construction. For $U>0$, double occupancy of a
trapping cell is forbidden, but a subset of these states remain exact
eigenstates: completely spin-polarized
states or states consisting of spatially sufficiently separated localized
electron excitations
(like the two bold valleys in Fig.~\ref{fig:sawtooth})
remain exact eigenstates also for $U>0$. Positivity
of the Coulomb term in the Hubbard model (\ref{eq:Hub}) implies that
such states remain in fact {\em ground states} for $U > 0$.
In fact, the same class of states are also exact eigenstates of the
$t-J$ model (\ref{eq:tJ}). However, in this case the magnetic interaction
term is no longer a positive operator. Thus, for sufficiently strong
$J>0$ other states may acquire lower energy, as has been observed
for the sawtooth chain with $J=2\,t$, $t'=\sqrt{2}\,t$ \cite{sawtooth-tJ}.
Therefore, the exact eigenstates under discussion are
ground states of the $t-J$ model (\ref{eq:tJ})
generally only for sufficiently small $J$.

One can tune the energy of the flat band to zero by setting $\mu=\mu_0$
with a suitable $\mu_0$. For the sawtooth chain with $t'=\sqrt{2}\,t$,
Eq.\ (\ref{eq:sawDisp}) shows that $\varepsilon_{-}(k)=0$ for
$\mu = \mu_0 = 2\,t$. All localized-electron GSs have zero
energy for $\mu = \mu_0$. It was argued in \cite{honecker_richter}
that the GS degeneracy is always macroscopic by giving an explicit lower
bound.
An alternative lower bound can be derived
along the lines of section 2.6.4 of \cite{RSH04}:
take a unit cell of the lattice with $l$ sites such that
it contains at least one trapping cell which does not overlap with the
corresponding trapping cell in the neighboring unit cells
(for the sawtooth chain one can use a unit cell of two valleys,
{\it i.e.}, $l=4$). Then each of these
trapping cells can be either empty or independently occupied by a
spin-up or -down electron. This yields $3^{N/l}$ GSs, {\it i.e.},
a lower bound for the GS entropy per site $S/N \ge \ln3/l$.

Two cases should be distinguished for the further analysis. If all
trapping cells are non-overlapping, spatial separation is automatically
ensured and each trapping cell can be
independently occupied by up to one electron in the
presence of repulsive interactions.
It is straightforward to count the
localized-electron GSs and compute their contribution
to thermodynamic quantities for any model of this type
\cite{monomer08}. Furthermore,
the average over the localized-electron GSs does not yield
any macroscopic magnetic moment in this case \cite{monomer08}.

In the case where trapping cells overlap, two electrons with different
spin projection generally feel the repulsive interaction if they are
localized in overlapping regimes. This favors parallel spin alignment
in the GS for larger electron fillings. If all trapping cells
are connected and each of them is occupied by one electron, these
arguments give rise to a fully saturated ferromagnetic GS
(flat-band ferromagnetism) \cite{mielke,tasaki,tasaki98}.
In order to count all localized-electron GSs, we
decompose the system into $n$-electron clusters, each of which
has a $n+1$-fold spin degeneracy \cite{sawtooth07}. In one dimension,
a suitable labeling yields a mapping to a dimer-type model which can be
solved with a transfer matrix method \cite{sawtooth07}. For the
sawtooth chain with $t'=\sqrt{2}\,t$ we find
a GS 
entropy per site $S/N=\ln((1+\sqrt{5})/2) = 0.48121\ldots$ \cite{sawtooth07},
which should be compared to the above lower bound
$S/N \ge \ln3/4=0.27465\ldots$
and the lower bound of \cite{honecker_richter}:
$S/N \ge \ln2/2=0.34657\ldots$. Note that the dimer mapping can be
applied to other one-dimensional models like
a model originally due to Watanabe \cite{IKWO98} or
two kagome-like chains \cite{jump}.

We emphasize that the localized-electron
states should form a basis of the GS manifold for $U=0$ in order
to ensure completeness also for $U>0$. This completeness condition can
be satisfied if the flat band is separated by a finite gap from the next
dispersive band, as is the case for the sawtooth chain. However,
if a dispersive band touches the flat band, this gives rise to additional
GSs.

\begin{figure}[!t]
\begin{center}
\includegraphics[width=\columnwidth]{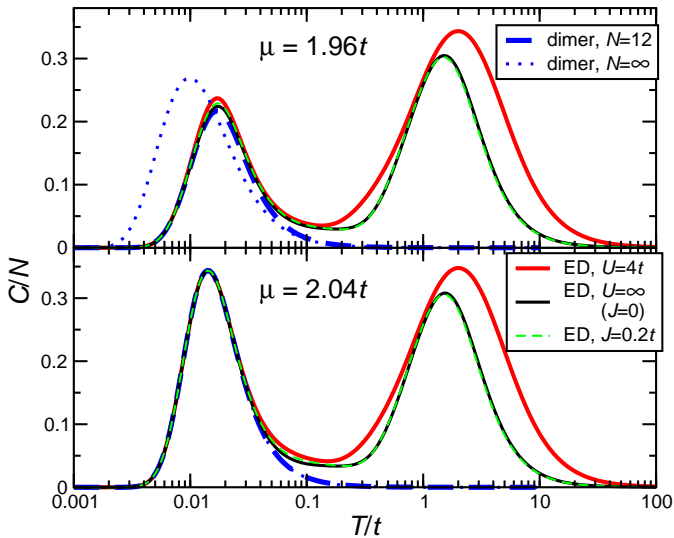}
\end{center}
\caption
{Specific heat $C$ per site $N$
in the grand canonical ensemble 
for the sawtooth chain with $t' = \sqrt{2}\,t$
for two values of the chemical potential $\mu$.
Exact diagonalization (ED) results are for $N=12$ sites.}
\label{fig:CmuLog}
\end{figure}

\section{Numerical results for the sawtooth chain}

Now we illustrate and extend some of the above general considerations
by presenting a comparison with exact diagonalization (ED) for
the sawtooth chain. 

Fig.~\ref{fig:CmuLog} shows the specific heat $C$ in the
{\em grand canonical ensemble} for $t' = \sqrt{2}\,t$ and
two values of $\mu$.
Firstly, we observe a low-temperature maximum around $T={\cal O}(10^{-2}t)$
in the ED results
for both values of the chemical potential. This low-temperature
maximum is (almost) the same for a finite-$U$ Hubbard model,
the $U=\infty$ Hubbard model (which is equivalent to the $J=0$ case of the
$t-J$ model), and even the $t-J$ model with $J=0.2\,t$.
This maximum is well described by the effective
dimer model \cite{sawtooth07}, {\it i.e.}, it arises from the
GS manifold which is split due to the deviation
$\mu \ne \mu_0 = 2\,t$. Comparison with other system sizes than the
$N=12$ results of Fig.~\ref{fig:CmuLog} reveals some
finite-size effects for $\mu < \mu_0$ while they are negligible
for $\mu > \mu_0$ \cite{sawtooth07}. Note that the limit $N \to \infty$
can be carried out in the effective
dimer model (dotted line in Fig.~\ref{fig:CmuLog}) \cite{sawtooth07}.
The remaining states give rise to another maximum in $C$
at a higher temperature $T={\cal O}(t)$. The area under this second
maximum is clearly smaller in the $t-J$ model than in the Hubbard model,
reflecting the reduced Hilbert space. Finite-size effects are
negligible in the high-temperature region \cite{sawtooth07}.

\begin{figure}[!t]
\begin{center}
\includegraphics[width=\columnwidth]{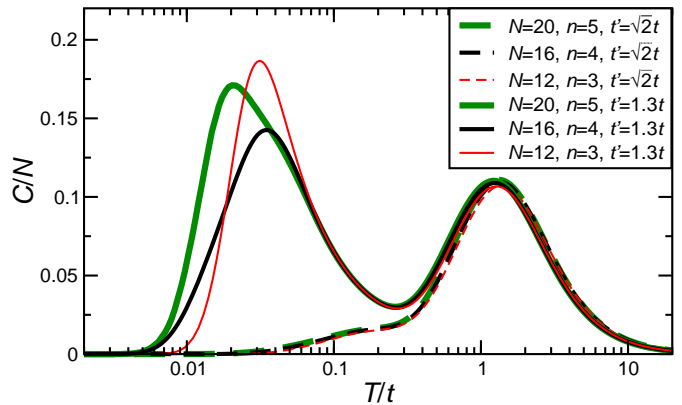}
\end{center}
\caption
{Specific heat $C$ per site $N$ in the canonical ensemble 
for the sawtooth chain with $n=N/4$ electrons.
All results are for the $U=\infty$ Hubbard model,
or equivalently the $t-J$ model with $J=0$.}
\label{fig:CnLog}
\end{figure}

We use the $t-J$ model with $J=0$ to discuss some complementary
aspects. The specific heat $C$ of the effective dimer
model vanishes identically in the {\em canonical ensemble}
since the energy of a localized electron state depends only on the electron
number $n$. Accordingly, the low-temperature maximum
in $C$ disappears 
for $t'=\sqrt{2}\,t$,
as illustrated by the ED results for $n=N/4$ shown in Fig.~\ref{fig:CnLog}.
However, a detuning $t' \ne \sqrt{2}\,t$ leads to a splitting of
the GS manifold such that the low-temperature maximum
reappears, as demonstrated in Fig.~\ref{fig:CnLog} for $t' = 1.3\,t$
(note that finite-size effects tend to be bigger in the canonical
ensemble than in the grand canonical ensemble).
This shows that the low-temperature maximum in $C$ is not only
robust under a small violation of the flat-band condition, but
that such a detuning actually helps to stabilize this feature.

\begin{figure}[!t]
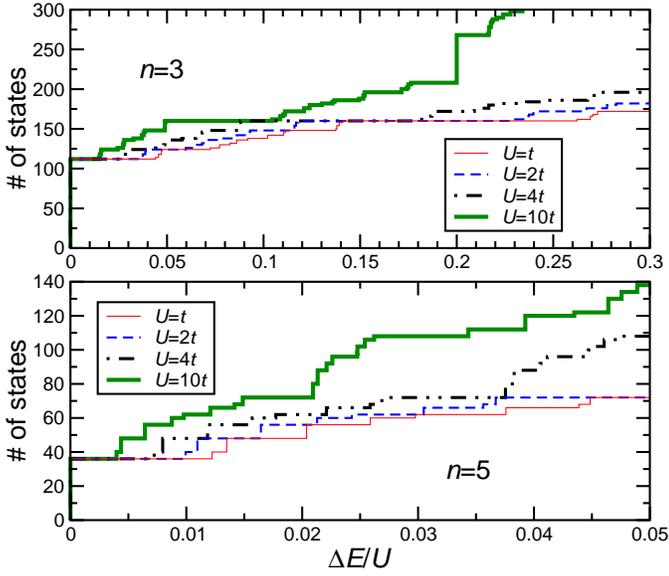

\begin{center}
\includegraphics[width=\columnwidth]{StatesN12n3.eps}
\includegraphics[width=\columnwidth]{StatesN12n5.eps}
\end{center}
\caption
{Integrated number of states with an excitation energy below $\Delta E$
for the sawtooth Hubbard model with $t' = \sqrt{2}\,t$, $N=12$ sites, and
different values of $U$.
The upper and lower panels show the sectors with $n=3$ and $5$ electrons,
respectively.}
\label{fig:StatesN12}
\end{figure}

\begin{figure}[!tb]
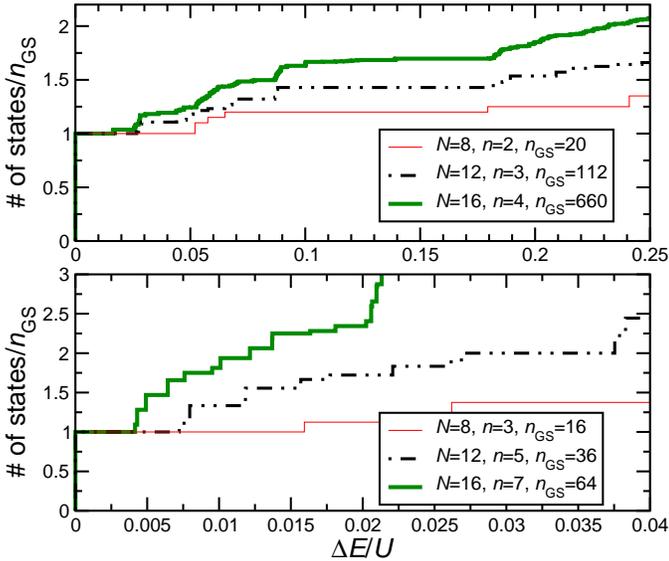

\begin{center}
\includegraphics[width=\columnwidth]{StatesU4No4.eps}
\includegraphics[width=\columnwidth]{StatesU4No2-1.eps}
\end{center}
\caption
{Integrated number of states 
normalized by the number of ground states $n_{\rm GS}$
for the sawtooth Hubbard model with $t' = \sqrt{2}\,t$ and $U=4\,t$.
The upper and lower panels show the sectors with $n=N/4$ and $n=N/2-1$ electrons,
respectively.}
\label{fig:StatesU4}
\end{figure}

The separation of the GS manifold at $t'=\sqrt{2}\,t$ from
states with $n > N/2$ is controlled by the charge gap $\Delta \mu$
which in the Hubbard model opens linearly as
$\Delta \mu \approx 0.46\,U$ for small $U \ll t$ \cite{sawtooth07}
and saturates at $\Delta \mu \approx 2\,t$ for large $U \gg t$
\cite{sawtooth07,cmt31}. We quantify the excitations in the sectors
with $n \le N/2$ using the integrated number of states with
an energy of at most $\Delta E$ above the degenerate GS
manifold. This quantity is shown in Fig.~\ref{fig:StatesN12}
for the $N=12$ Hubbard model with different values of $U$.
The case $n=N/4=3$ is representative of the generic situation while
the energy of the lowest excited state is smallest in the sector
$n=N/2-1=5$. We observe that an appreciable density of states appears
at rather low energies above the highly degenerate GSs.
The fact that the curves in Fig.~\ref{fig:StatesN12} are very similar
for small $U$ when $\Delta E$ is scaled by $U$ indicates that these
low-lying excitations originate from
states which used to be GSs for $U=0$.
The finite-size dependence at $n=N/4$ and $n=N/2-1$
is analyzed in Fig.~\ref{fig:StatesU4} for the example $U=4\,t$.
In order to be able to compare different values of $N$,
we divide the integrated number of states by the number of
GSs $n_{\rm GS}$ (see legends of Fig.~\ref{fig:StatesU4}
for the values). We observe that the density of
low-energy excitations increases with $N$, even when compared to $n_{\rm GS}$.
It is not completely clear at present whether this indicates
the absence of a thermodynamic excitation gap for $N\to\infty$,
which would imply quantitative corrections to the dimer model for
all temperatures $T>0$. In any case, this large density of
low-lying excitations is probably the origin of the small
deviations for $\mu = 1.96\,t$ in Fig.~\ref{fig:CmuLog}
between the effective dimer model and the Hubbard model
with $U=4\,t$.


\section{Conclusions}

We have argued that in highly frustrated Hubbard and $t-J$ models
one finds a macroscopic GS degeneracy for a certain
value of the chemical potential $\mu_0$ or, equivalently,
in a certain range of electron fillings. A splitting of this
GS manifold, e.g., by a small deviation $\mu \ne \mu_0$
or deviation from the ideal flat-band geometry leads to a
characteristic low-temperature peak in the specific heat.
These general considerations have been illustrated with
ED results for the sawtooth chain. We have also exhibited
a large number of low-lying excited states in the sawtooth
chain.

\section{Acknowledgments}

A.H.\ acknowledges financial support by the German Science Foundation
(DFG) through a Heisenberg fellowship (Project HO~2325/4-1).
We are grateful for allocation of CPU time at the HLRN Hannover.
Part of the computations for the $t-J$ model have been
carried out with the ALPS fulldiag application \cite{ALPS}.

\end{document}